\documentclass[aps,prd,twocolumn]{revtex4-2}

\usepackage{amssymb}
\usepackage{amsmath}

\usepackage{graphics}
\usepackage[export]{adjustbox}

\usepackage{xparse}

\usepackage{hyperref}
\usepackage{hypernat}

\usepackage{layouts}

\usepackage[caption=false,labelfont=normalsize,labelformat=empty, textfont=normalsize,justification=centering]{subfig}

\renewcommand{\Im}{\mathrm{Im}}

\newcommand{\iu}{\mathrm{i}\hskip0.07em}
\newcommand{\eq}{\mathrm{eq}}

\begin{document}

\title{Unitarity, real-intermediate states, and fixed-order approach to resonant dark matter annihilation}

\author{Peter Mat\'ak}
\email{peter.matak@fmph.uniba.sk}
\affiliation{Department of Theoretical Physics, Comenius University,\\ Mlynsk\'a dolina, 84248 Bratislava, Slovak Republic}

\date{\today}

\begin{abstract}
We study the role of perturbative unitarity in the resonant annihilation of two dark matter particles into the standard model bath. Systematically including all kinematically allowed holomorphic cuts of the corresponding forward-scattering diagram, cancelation of the singularities occurs, resulting in a fixed-order correction to the narrow-width approximation for the annihilation cross section. Unlike the standard approach based on including the finite width of the mediator, no double counting of intermediate states occurs.
\end{abstract}

\maketitle

\section{Introduction}\label{sec1}

Among dark matter production mechanisms, one of the known exceptions from the standard relic density calculation originates from the resonant enhancement of the thermally averaged annihilation cross section \cite{Griest:1990kh, Gondolo:1990dk}. Resonance typically appears in the interactions mediated by $s$-channel exchanges of unstable particles heavier than two dark matter masses. Various aspects of this scenario, including the effect of kinetic decoupling \cite{Duch:2017nbe, Binder:2017rgn, Ala-Mattinen:2019mpa, Binder:2021bmg}, gauge invariance \cite{Duch:2018ucs}, thermal corrections \cite{Kim:2016kxt}, and higher-order perturbative corrections \cite{Laine:2022ner}, have been studied in the literature in recent years. In this paper, we focus on the role of perturbative unitarity in treating unstable particle intermediate states. 

To avoid the singularity when the centre-of-mass energy approaches the mediator mass, the resonant propagator is usually considered with a finite imaginary part. Then the narrow-width approximation may be employed, which can be understood as an on-shell production of the mediator with the cross section multiplied by the branching ratio of its decay into standard model particles \footnote{The accuracy of the narrow-width approximation has been studied in Refs. \cite{Berdine:2007uv, Kauer:2007zc, Uhlemann:2008pm}. Our approach goes beyond the narrow-width approximation by including the off-shell mediator exchanges.}. While the former originates from the summing of all perturbative orders of the Dyson series, the latter corresponds to the leading-order calculation. Following the previous work in Ref. \cite{Matak:2022qwc}, our approach is in-between the two possibilities, allowing for a next-to-leading-order correction without double counting the leading-order result \footnote{A detailed discussion of the double counting problem and various subtraction schemes has recenlty been presented in Ref. \cite{Ala-Mattinen:2023rbm}}. Here, instead of Cutkosky cuts \cite{Cutkosky:1960sp}, we use holomorphic cutting rules \cite{Coster:1970jy, Bourjaily:2020wvq, Blazek:2021olf, Hannesdottir:2022bmo}, which are more convenient for tracking real intermediate states. Furthermore, at one loop, the results are equivalent to what would be obtained within the modified perturbation theory \cite{Tkachov:1997aq, Tkachov:1998uy, Nekrasov:2007ta}, which, to our knowledge, has not been applied to resonant dark matter annihilation.

We further employ a model in which, for simplicity, all particles are real scalars. The dark matter particles $\chi$ of mass $m$ annihilate into a pair of massless $\varphi$ particles, while the mediator $\phi$ of mass $M$ and width $\Gamma$ couples to both $\chi$ and $\varphi$ fields through the Lagrangian density
\begin{align}\label{eq1}
\mathcal{L}=-\frac{1}{2}\lambda_{\chi}\phi\chi^2-\frac{1}{2}\lambda_{\varphi}\phi\varphi^2.
\end{align}
Modifying the calculations presented here to more complicated models is straightforward. 

The rest of the paper is structured as follows. In section \ref{sec2}, we apply the holomorphic cutting rules to a simple $s$-channel diagram for dark matter annihilation, obtaining a nonsingular result. In section \ref{sec3}, the connection between the fixed-order approach and the Breit-Wigner approximation is discussed. Finally, in section \ref{sec4}, the accuracy of our calculation for predicting the dark matter abundance is studied.

\section{Holomorphic cutting rules and real-intermediate states}\label{sec2}

Concerning singularities in particle physics, their presence often indicates incompleteness in the calculation. As discussed in Refs. \cite{Racker:2018tzw, Blazek:2021gmw, Blazek:2022azr}, forgetting anomalous thresholds in right-handed neutrino interactions with quarks leaves uncancelled infrared divergences. Their inclusion, on the other hand, leads to a finite result by the Kinoshita-Lee-Nauenberg theorem \cite{Kinoshita:1962ur, Lee:1964is, Frye:2018xjj}. 

Unitarity is thus a good starting point when looking for a complete set of contributions that may potentially lead to a non-singular outcome. For any initial state of interest, this can generally be achieved by constructing all forward-scattering diagrams of desired perturbative order and cutting them in all kinematically allowed ways with complex conjugation at one side of the cut.
The $S$-matrix unitarity with $\iu T=S-1$ implies
\begin{align}\label{eq2}
2\Im T^{\vphantom{\dagger}}_{ii} = \sum_f \vert T^{\vphantom{\dagger}}_{fi}\vert^2.
\end{align}
It has been argued in Refs. \cite{Coster:1970jy, Bourjaily:2020wvq, Blazek:2021olf, Hannesdottir:2022bmo} that in the above equation, the right-hand side can be replaced by
\begin{align}\label{eq3}
-\sum_f\iu T^{\vphantom{\dagger}}_{if}\iu T^{\vphantom{\dagger}}_{fi}+\sum_{f,k}\iu T^{\vphantom{\dagger}}_{\vphantom{f}ik}\iu T^{\vphantom{\dagger}}_{kf} \iu T^{\vphantom{\dagger}}_{fi} - \ldots
\end{align}
and instead of a single cut with complex conjugation, each forward-scattering diagram has to be cut, with alternating signs, as many times as allowed by the kinematics \footnote{It is the absence of the complex conjugation why this modification of cutting rules has been named holomorphic in Ref. \cite{Hannesdottir:2022bmo}.}. The cut diagrams will be understood further as contributions to $\vert M_{fi}\vert^2=\vert T_{fi}\vert^2/V_4$ with $V_4=(2\pi)^4\delta^{(4)}(0)$. Integration over the final-state momenta in the cut lines will be implicit.

If an uncancelled singularity occurs, we often have to look at the right-hand side of Eq. \eqref{eq2} and check if all possible final states $f$, i.e. all cuts in the respective diagrams, have been properly included. It remarkably turns out that the singularity coming from the integration over $s$ in the square of a tree-level amplitude containing
\begin{align}\label{eq4}
\left\vert\frac{1}{s-M^2+\iu\epsilon} \right\vert^2
\end{align}
can be treated in a very similar way. To see that, let us consider the forward-scattering diagram, cutting which contributes to the $s$-channel-annihilation amplitude squared. In fact, there are three independent cuts that can be made simultaneously. Applying the holomorphic cutting rules of Eq. \eqref{eq3} yields
\begin{align}\label{eq5}
&-\hskip1mm\includegraphics[scale=1,valign=c]{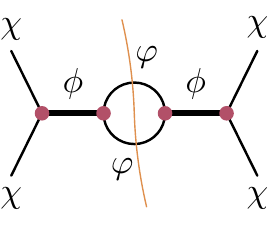}\hskip1mm
-\hskip1mm\includegraphics[scale=1,valign=c]{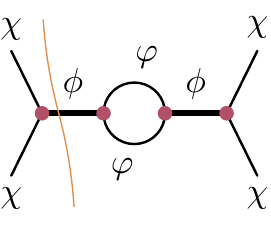}\\
&-\hskip1mm\includegraphics[scale=1,valign=c]{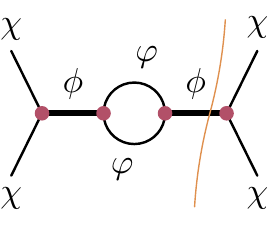}\hskip1mm
+\hskip1mm\includegraphics[scale=1,valign=c]{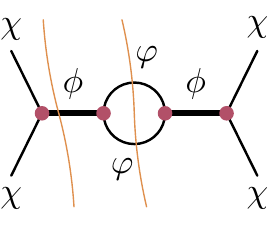}\nonumber\\
&+\hskip1mm\includegraphics[scale=1,valign=c]{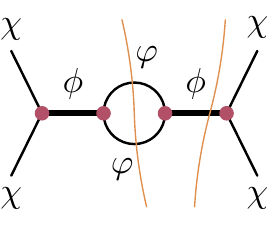}\hskip1mm
+\hskip1mm\includegraphics[scale=1,valign=c]{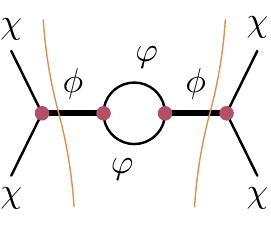}\nonumber\\
&-\hskip1mm\includegraphics[scale=1,valign=c]{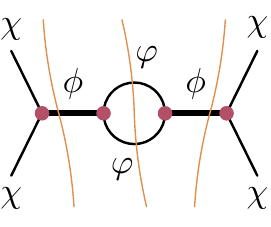}\nonumber
\end{align}
where the cut diagrams contain singular expressions and should be treated as distributions. 

Except for the last term in Eq. \eqref{eq5}, each diagram contains an uncut subdiagram indicating the presence of on-shell intermediate states or, by unitarity, a nonvanishing imaginary part. Hence, for the uncut $\phi$ propagators, we employ
\begin{align}\label{eq6}
\frac{1}{s-M^2+\iu\epsilon} = \mathcal{P}\frac{1}{s-M^2}-\iu\pi\delta(s-M^2)
\end{align}
or diagrammatically
\begin{align}\label{eq7}
\includegraphics[scale=1,valign=c]{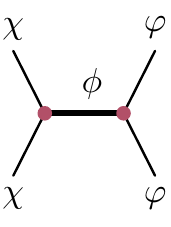}\hskip1mm = 
\hskip1mm\includegraphics[scale=1,valign=c]{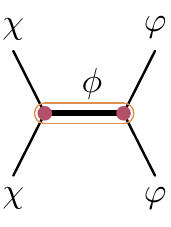}\hskip1mm +
\frac{1}{2}\hskip1mm\includegraphics[scale=1,valign=c]{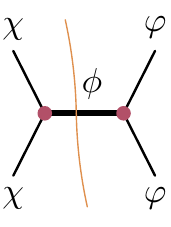}
\end{align}
where the encircled line stands for the principal value or the propagator's real part. Similarly, the self-energy loop can be split into real and imaginary parts as
\begin{align}\label{eq8}
\includegraphics[scale=1,valign=c]{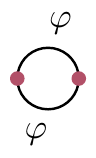}\hskip1mm = 
\hskip1mm\includegraphics[scale=1,valign=c]{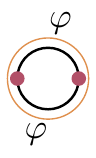}\hskip1mm +
\frac{1}{2}\hskip1mm\includegraphics[scale=1,valign=c]{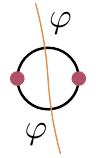}.
\end{align}
Plugging Eqs. \eqref{eq7} and \eqref{eq8} in Eq. \eqref{eq5}, the on-shell intermediate states are made explicit, leading to several cancelations, and we are finally left with two non-singular pieces. The first of them corresponds to
\begin{align}\label{eq9}
-\includegraphics[scale=1,valign=c]{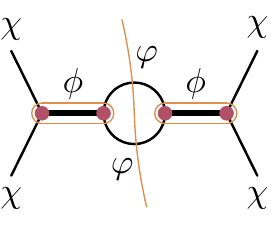}\hskip1mm
-\frac{1}{4}\hskip1mm\includegraphics[scale=1,valign=c]{math1g.pdf}
\end{align}
where the principal value and the delta function, both squared, combine into
\begin{align}\label{eq10}
\frac{(s-M^2)^2-\epsilon^2}{[(s-M^2)^2+\epsilon^2]^2}=
-\frac{\partial}{\partial s}\mathcal{P}\frac{1}{s-M^2}
\end{align}
which is integrable when it comes to thermal averaging. Unlike the regularisation by the finite mediator width, Eq. \eqref{eq9} allows us to properly account for off-shell mediator exchanges without double counting the leading-order result \footnote{See the discussion near Eq. (13) in Ref. \cite{Matak:2022qwc}.}. Furthermore, double counting is avoided without employing complex-valued counterterms, as it is in the complex mass scheme \cite{Denner:2014zga}. Therefore, despite common wisdom, the summation of the Dyson series is unnecessary unless the resonance occurs very close to the on-shell production threshold.

The leading-order contribution originates from 
\begin{align}\label{eq11}
-\hskip1mm\includegraphics[scale=1,valign=c]{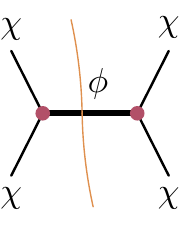}\hskip1mm
\times\mathcal{B}(\phi\rightarrow\varphi\varphi)
\end{align}
corresponding to the $\chi\chi\rightarrow\phi$ inverse decay, which is multiplied by the branching ratio
\begin{align}\label{eq12}
\mathcal{B}(\phi\rightarrow\varphi\varphi)=
\includegraphics[scale=1,valign=c]{math4c.pdf}\Bigg/\Bigg\{
\includegraphics[scale=1,valign=c]{math4c.pdf}\hskip1mm+\hskip1mm
\includegraphics[scale=1,valign=c]{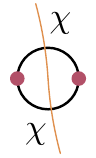}\Bigg\}
\end{align}
to obtain the contribution to the annihilation of dark matter.

In addition to Eq. \eqref{eq9}, we further obtain from Eq. \eqref{eq5} two terms that contain the real part of the inserted loop
\begin{align}\label{eq13}
-\hskip1mm\includegraphics[scale=1,valign=c]{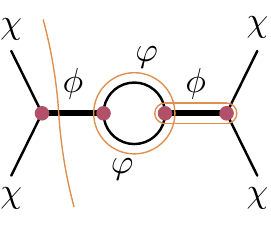}\hskip1mm
-\hskip1mm\includegraphics[scale=1,valign=c]{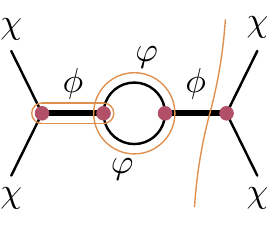}
\end{align}
where the product of the principal value and the delta function gives
\begin{align}\label{eq14}
2\pi\delta(s-M^2)\mathcal{P}\frac{1}{s-M^2}=-\frac{\partial}{\partial s}\pi\delta\big(s-M^2\big).
\end{align}
As indicated by the cuts in Eq. \eqref{eq13}, these diagrams represent a higher-order correction to the $\chi\chi\rightarrow\phi$ inverse decay and again have to be multiplied by the branching ratio of Eq. \eqref{eq12}. Integrating the derivative in Eq. \eqref{eq14} by parts results in terms that contain the self-energy and its derivative evaluated at $s=M^2$. These represent the mass shift and the Lehmann-Symanzik-Zimmermann reduction factor of the $\phi$ external leg, respectively \cite{Matak:2022qwc}.

\section{Unitarity and Breit-Wigner approximation}\label{sec3}

In this section, we briefly comment on the role of unitarity in the standard treatment based on the summation of the Dyson series. Following the procedure introduced in Ref. \cite{Matak:2022qwc}, we complete the one-loop self-energy of $\phi$  as
\begin{align}\label{eq15}
-\iu\Sigma(s) = \includegraphics[scale=1,valign=c]{math4a.pdf}+
\includegraphics[scale=1,valign=c]{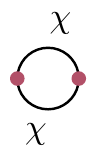}
\end{align}
defining $\Sigma_R$ and $\Sigma_I$ as the real and imaginary parts, respectively. Then the expression in Eq. \eqref{eq9} can be rewritten as
\begin{align}\label{eq16}
2\lambda^2_\chi\Sigma_I(s)\bigg\{\frac{\partial}{\partial s}\mathcal{P}\frac{1}{s-M^2}\bigg\}\times\mathcal{B}(\phi\rightarrow\varphi\varphi)
\end{align}
where the imaginary part of the second loop added in Eq. \eqref{eq15} is cancelled by the denominator in the branching ratio of Eq. \eqref{eq12}.

The situation is different for the contribution in Eq. \eqref{eq13}, which has to be extended as
\begin{align}\label{eq17}
2\lambda^2_\chi\Sigma_R(s)\bigg\{-\frac{\partial}{\partial s}\pi\delta\big(s-M^2\big)\bigg\}
\times\mathcal{B}(\phi\rightarrow\varphi\varphi)
\end{align}
to include the real part of both loops in Eq. \eqref{eq15}. We further employ the renormalisation conditions for which $\Sigma_R$ and its derivative vanish at $s=M^2$. Then the contribution of Eq. \eqref{eq17} vanishes as well, and the only next-to-leading-order correction comes from Eq. \eqref{eq16}. 
 
After summing the Dyson series, the finite width in the propagator results from the nonvanishing imaginary part of the self-energy. Therefore, instead of a single loop insertion as in Eq. \eqref{eq5}, we can sum over any number of loops inserted in the propagator of $\phi$. Then, we can apply the holomorphic cutting rules to each forward-scattering diagram obtained in this way, as in Ref. \cite{Hannesdottir:2022bmo}. Alternatively, we can employ Eq. \eqref{eq16} and \eqref{eq17} with the one-particle-irreducible self-energy replaced by
\begin{align}\label{eq18}
-\iu\Tilde{\Sigma}(s)=-\iu\Sigma(s) \times\sum_{n=0}^\infty\left(\frac{\Sigma(s)}{s-M^2+\iu\epsilon}\right)^n
\end{align}
leading to \cite{Matak:2022qwc}
\begin{align}
\Tilde{\Sigma}_R&=\frac{(s-M^2)^2\Sigma_R-(s-M^2)(\Sigma^2_R+\Sigma^2_I)}{(s-M^2-\Sigma_R)^2+\Sigma^2_I}\label{eq19}\\
\Tilde{\Sigma}_I&=\frac{(s-M^2)^2\Sigma_I}{(s-M^2-\Sigma_R)^2+\Sigma^2_I}\label{eq20}
\end{align}
where we put $\epsilon=0$ after the summation. From Eq. \eqref{eq19} we can immediately see that
\begin{align}\label{eq21}
\Tilde{\Sigma}_R(s)\bigg\vert_{s=M^2}=0,\quad\frac{\partial\Tilde{\Sigma}_R(s)}{\partial s}\bigg\vert_{s=M^2}=-1.
\end{align}
When replacing $\Sigma_R$ by $\Tilde{\Sigma}_R$ in Eq. \eqref{eq17}, the only non-vanishing part contains its derivative and cancels the leading-order contribution of Eq. \eqref{eq11}. This result agrees with Ref. \cite{Veltman:1963th} stating no unstable particles need to be included in the initial or final states \footnote{See also the discussion on page 23 in Ref. \cite{Hannesdottir:2022bmo}.}. Instead, we should only include them in resummed propagators, as seen from Eqs. \eqref{eq16} and \eqref{eq20} combined into
\begin{align}\label{eq22}
\frac{-2\lambda^2_\chi\Sigma_I(s)}{(s-M^2-\Sigma_R(s))^2+\Sigma_I(s)^2}
\times\mathcal{B}(\phi\rightarrow\varphi\varphi).
\end{align}
Neglecting the energy dependence of the self-energy after the appropriate counterterms are included in its real part, the usual Breit-Wigner approximation is obtained.

\section{Thermal averaging and dark matter relic density}\label{sec4}

\begin{figure*}
\subfloat{\label{fig1a}}
\subfloat{\label{fig1b}}
\centering\includegraphics{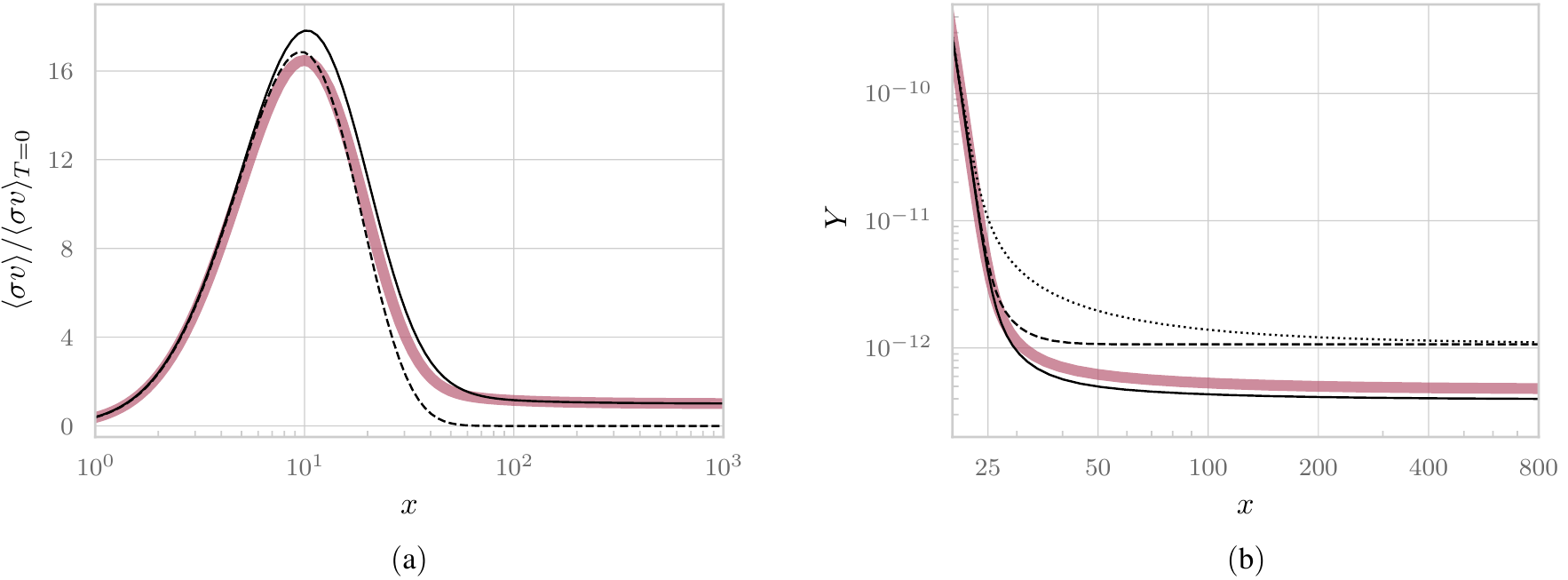}
\caption{\label{fig1} Thermally averaged annihilation cross section as a function of $x$ at the leading order (dashed black line), combined leading and next-to-leading order (solid black line), and the Breit-Wigner approximation (thick red line) using $M=100~\mathrm{TeV}$, $\Delta = -0.2$, $\gamma=0.01$, and $\alpha=0.1$ is shown in Fig. \ref{fig1a}. In Fig. \ref{fig1b}, the evolution of the dark matter number density over the entropy density is plotted for the same cases. The dotted line corresponds to the zero-temperature cross section in Eq. \eqref{eq31}.}
\end{figure*}

To study the accuracy of the fixed-order approximation in the calculation of the dark matter relic abundance, we need to solve the Boltzmann equation for the number density to entropy density ratio \cite{Gondolo:1990dk}
\begin{align}\label{eq23}
\frac{dY}{dx}=-\bigg(\frac{45}{\pi}G\bigg)^{-1/2}\frac{g^{1/2}_{*}m}{x^2}
\langle\sigma v\rangle\big(Y^2-Y^2_\mathrm{eq}\big).
\end{align}
In this equation, $x=m/T$, $G$ is the gravitational constant, and $Y_\mathrm{eq}=n_\mathrm{eq}/s$ with
\begin{align}\label{eq24}
n_\mathrm{eq}(x)=\frac{m^3}{2\pi^2}\frac{K_2(x)}{x}\quad\text{and}\quad
s(x)=\frac{2\pi^2}{45}g_{*}\frac{m^3}{x^3}
\end{align}
where $K_2(x)$ is the modified Bessel function of the second kind. The number of effective degrees of freedom is set at $g_{*}=106.75$ as if all particles in the standard model were ultrarelativistic. 

It may seem that Eq. \eqref{eq23} only represents the dark matter density evolution due to two-particle annihilations. At the leading and next-to-leading order, we also deal with decays and inverse decays; thus, a system of two Boltzmann equations for $\chi$ and $\phi$ particles has to be considered. Although this statement is generally correct, we argue in the appendix that for the dark matter freeze-out, the single Eq. \eqref{eq23} can also account for decays and inverse decays with high precision.

Following the convention of Refs. \cite{Guo:2009aj, Ibe:2008ye}, we introduce 
\begin{align}\label{eq25}
\Delta = 1-\frac{M^2}{4m^2}<0\quad\text{and}\quad\gamma=\frac{\Gamma}{M}
\end{align}
while the annihilation cross section is proportional to the square of $\alpha=\lambda_\chi\lambda_\varphi/(4M^2)$. These parameters are not independent for the particular model in Eq. \eqref{eq1}, we have to require
\begin{align}\label{eq26}
\frac{4\pi\gamma}{\alpha}\geq\bigg(\frac{-\Delta}{1-\Delta}\bigg)^{1/4}.
\end{align}
Furthermore, we focus only on $-\Delta\gg\gamma$. Otherwise, the summation of the Dyson series with energy-dependent self-energy may become unavoidable \cite{Duch:2017nbe}.

With the new parameterisation, the square of the leading-order amplitude in Eq. \eqref{eq11}, including the multiplication by the branching ratio, can be written as
\begin{align}\label{eq27}
\vert M\vert^2_\mathrm{LO}=\alpha^2\frac{1-\Delta}{\gamma}\delta(\Tilde{s}-1+\Delta)
\end{align}
where we introduce $\Tilde{s}=s/(4m^2)$. This allows us to write the respective thermally averaged cross section as \cite{Gondolo:1990dk}
\begin{align}\label{eq28}
\langle\sigma v\rangle_\mathrm{LO}&= \frac{1-\Delta}{M^2}\frac{2x}{K_2(x)^2}
\int\displaylimits^\infty_1 d\Tilde{s}\sqrt{\Tilde{s}-1}K_1(2x\sqrt{\Tilde{s}})
\vert M\vert^2_\mathrm{LO}\nonumber\\
&=\alpha^2\frac{\sqrt{-\Delta}}{\gamma}\frac{(1-\Delta)^2}{M^2}\frac{2xK_1(2x\sqrt{1-\Delta})}{K_2(x)^2}.
\end{align}

To account for higher-order corrections, we need to add to the above expression a contribution in which $\vert M\vert^2_\mathrm{LO}$ in the first row of Eq. \eqref{eq28} is replaced by 
\begin{align}\label{eq29}
\vert M\vert^2_{\mathrm{NLO}}=&\frac{\alpha^2}{\pi}(1-\Delta)^2\bigg\{-\frac{\partial}{\partial\Tilde{s}}\mathcal{P}\frac{1}{\Tilde{s}-1+\Delta}\bigg\}
\end{align}
from Eq. \eqref{eq16}, obtaining $\langle\sigma v\rangle_{\mathrm{LO}+\mathrm{NLO}}$. It is important to emphasise that the total cross section is still a distribution for the squared amplitude in Eq. \eqref{eq29}. In other words, $\epsilon$ in Eq. \eqref{eq10} can not be set to zero until integration over $\Tilde{s}$ is performed. 

Finally, we compute the cross-section average for the Breit-Wigner approximation, denoted $\langle\sigma v\rangle_{\mathrm{BW}}$, using
\begin{align}\label{eq30}
\vert M\vert^2_{\mathrm{BW}}=&\frac{\alpha^2}{\pi}
\frac{1}{[\Tilde{s}/(1-\Delta)-1]^2+\gamma^2}
\end{align}
from which the leading-order result in Eq. \eqref{eq27} can be obtained in the narrow-width approximation. For a fermion dark matter and vector mediator, a similar expression has been considered in Ref. \cite{Binder:2021bmg}.

\begin{figure*}
\subfloat{\label{fig2a}}
\subfloat{\label{fig2b}}
\centering\includegraphics{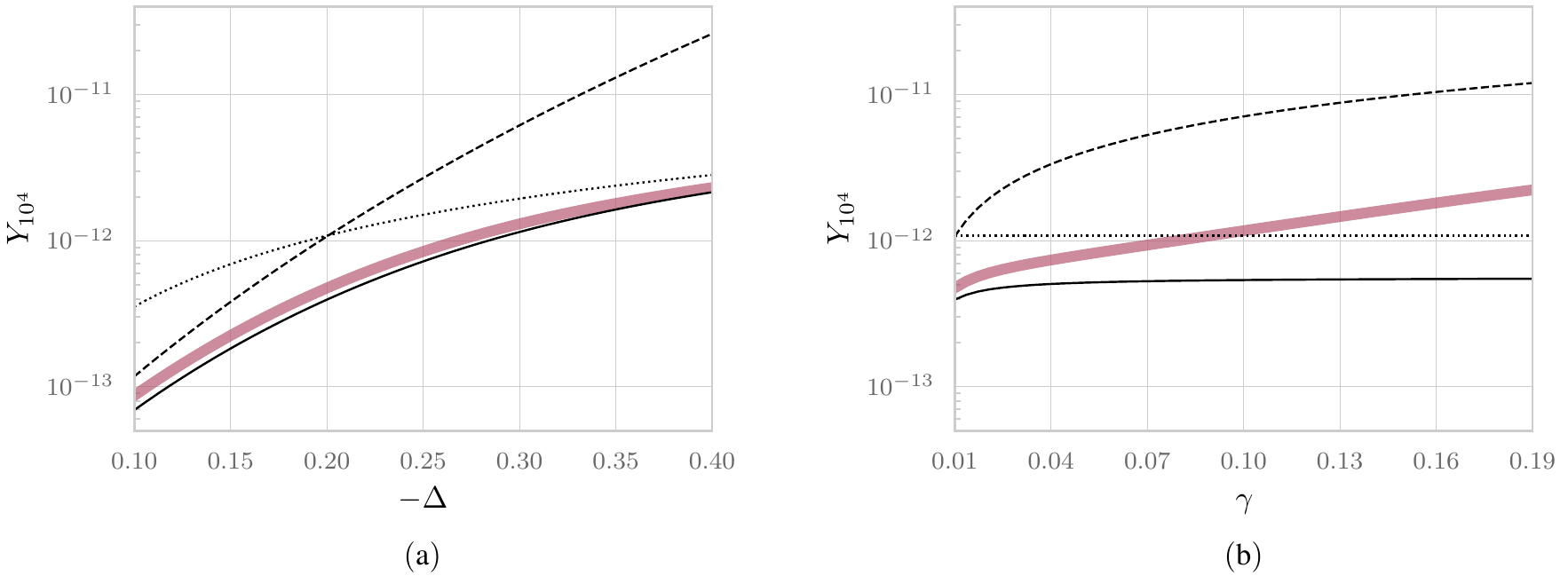}
\caption{\label{fig2} Dark matter relic density over the entropy density for $M=100~\mathrm{TeV}$ and $\alpha=0.1$ calculated at $x=10^4$ as a function of $\Delta$ with $\gamma=0.01$ (Fig. \ref{fig2a}) or as a function of $\gamma$ with $\Delta=-0.2$ (Fig. \ref{fig2b}). The lines correspond to different cross section approximations as in Fig. \ref{fig1}.}
\end{figure*}

We plot the thermally averaged cross section in Fig. \ref{fig1a}. At low temperatures, such that $x>100$, both $\langle\sigma v\rangle_{\mathrm{LO}+\mathrm{NLO}}$ and $\langle\sigma v\rangle_{\mathrm{BW}}$ approach the zero-temperature limit
\begin{align}\label{eq31}
\langle\sigma v\rangle_{T=0} = \frac{\alpha^2}{\pi}\frac{(1-\Delta)^3}{M^2\Delta^2}.
\end{align} 
Moreover, as seen in Fig. \ref{fig1b}, the dark matter density evolution obtained from Eq. \eqref{eq30} is relatively well approximated by including the next-to-leading-order correction of Eq. \eqref{eq29}. The zero-temperature and narrow-width approximations fail almost by an order of magnitude in the final relic density.

The accuracy of the fixed-order calculation also depends on how close the dark matter mass is to half of the mediator mass. The relic density as a function of $-\Delta$ is shown in Fig. \ref{fig2a}. For large values of $-\Delta$, the off-shell mediator exchanges are important, and the leading-order narrow-width approximation fails. For smaller $-\Delta$ values, on the other hand, the narrow width approximation may be accurate if $-\Delta\gg\gamma$. Then, the zero-temperature approximation becomes insufficient, as it does not include the resonance. Our next-to-leading-order calculation smoothly interpolates the two limiting cases for the range of values considered. 

In Fig. \ref{fig2b}, we plot the relic density over the entropy density as a function of $\gamma$. The accuracy of the next-to-leading-order calculation decreases as $\gamma$ approaches the value of $-\Delta$. In that case, the near-threshold region is more important, and the summation of the Dyson series becomes necessary.

Before we summarise, let us comment on the relation of the present work to the existing literature. In Ref. \cite{Laine:2022ner}, thermal field theory has been used to derive similar fixed-order approximations within the scalar singlet dark matter model. At the leading order, the thermally averaged cross section in Eq. \eqref{eq28} is analogous to Eq. (B.7) in Ref. \cite{Laine:2022ner}. However, explicit multiplication by the branching ratio seems to have been omitted. In our opinion, this is a consequence of a specific choice of parameters due to the experimental constraints requiring a very weak dark matter coupling. The branching ratio of the Higgs boson decaying into standard-model particles is therefore approximately equal to unity. At the next-to-leading order, not all results were shown explicitly in Ref. \cite{Laine:2022ner}, but some of the features, such as regularisation of the second-order poles by principal value derivatives, are mentioned.

\section{Summary}\label{sec5}

In this work, we studied the accuracy of a unitarity-based perturbative approach to resonant dark matter annihilation. Using the holomorphic cutting rules, we derived an approximation for the off-shell mediator exchanges. These usually lead to a singularity in the $s$-channel diagram squared. Following our previous work in Ref. \cite{Matak:2022qwc}, we have shown that the singularity disappears when all contributions of the given perturbative order are properly included. Therefore, unless the resonance occurs near the on-shell production threshold, the summation of the Dyson series, producing a finite mediator width, is unnecessary.

We have solved the Boltzmann equation for the dark matter relic density numerically. The next-to-leading-order corrections, combined with the on-shell mediator production representing the leading-order result, were compared to the annihilation with the Breit-Wigner cross section.

\begin{acknowledgements}
The author was supported by the Slovak Slovak Education Ministry Contract No. 0466/2022 and by the Slovak Grant Agency VEGA, project No. 1/0719/23.
\end{acknowledgements}

\appendix*
\section{Boltzmann equations including decays and annihilations}

At the lowest order, the Boltzmann equations for the $\chi$ and $\phi$ particle densities receive contributions from decays and inverse decays. Throughout this work, we used the single Boltzmann equation for two-to-two processes only. In this appendix, we argue that it is an accurate approximation for the resonant annihilation freeze-out. 

To simplify the notation, let us denote the $\phi\rightarrow\varphi\varphi$ branching ratio in Eq. \eqref{eq12} as $\mathcal{B}$, the branching ratio of the $\phi\rightarrow\chi\chi$ decay as $1-\mathcal{B}$, and the leading-order thermally averaged decay width of $\phi$ as $\langle\Gamma\rangle$. Then, we can write
\begin{align}
xH\frac{dY_{\chi}}{dx}=2(1-\mathcal{B})\langle\Gamma\rangle\Bigg[
Y_\phi-\Bigg(\frac{Y_\chi}{Y_{\chi,\eq}}\Bigg)^2 Y_{\phi,\eq}\Bigg]\label{A1}\\
xH\frac{dY_{\phi}}{dx}= -(1-\mathcal{B})\langle\Gamma\rangle\Bigg[
Y_\phi-\Bigg(\frac{Y_\chi}{Y_{\chi,\eq}}\Bigg)^2 Y_{\phi,\eq}\Bigg]\label{A2}\\
-\mathcal{B}\langle\Gamma\rangle\Big(Y_\phi-Y_{\phi,\eq}\Big)\nonumber
\end{align}
where the factor of $2$ in Eq. \eqref{A1} compensates for the final-state symmetry factor in $\langle\Gamma\rangle$. Multiplying Eq. \eqref{A2} by $2(1-\mathcal{B})$ and adding it to the equation \eqref{A1}, we obtain
\begin{align}\label{A3}
xH\frac{dY_\chi}{dx}+2(1-\mathcal{B})xH\frac{dY_{\phi}}{dx}=\\
-2(1-\mathcal{B})\mathcal{B}\langle\Gamma\rangle Y_{\phi,\eq}\Bigg[\Bigg(\frac{Y_\chi}{Y_{\chi,\eq}}\Bigg)^2-1\Bigg].\nonumber
\end{align}
As $\phi$ is more than twice as heavy as $\chi$, if both these particle species were in thermal equilibrium until a relatively large value of $x$, we get 
\begin{align}\label{A4}
Y_{\phi}\simeq Y_{\chi}\exp\{-2x\sqrt{1-\Delta}+x\}
\end{align}
before the decoupling. Therefore, the abundance of $\phi$ is suppressed by several orders of magnitude compared to that of $\chi$. This suppression cannot be overcome by early decoupling of $\phi$, as both Eqs. \eqref{A1} and \eqref{A2} share the same $\phi\leftrightarrow\chi\chi$ part. We may thus introduce 
\begin{align}\label{A5}
Y=Y_\chi+2(1-\mathcal{B})Y_\phi  
\end{align}
and rewrite Eq. \eqref{A3} in terms of $Y\approx Y_\chi$, while neglecting contributions of the $\mathcal{O}(Y_\phi/Y_\chi)$ order on the right-hand side. The procedure results in Eq. \eqref{eq23} with thermally-averaged cross section written as
\begin{align}\label{A6}
\langle\sigma v\rangle_{\mathrm{LO}}=2(1-\mathcal{B})\mathcal{B}
\frac{\langle\Gamma\rangle Y_{\phi,\eq}}{s Y^2_{\chi,\eq}}
\end{align}
equal to the expression in Eq. \eqref{eq28}, including the multiplication by the branching ratio of Eq. \eqref{eq12}. 

At higher orders, annihilations through off-shell $\phi$ exchanges also contribute to the density evolution of $\chi$ particles. However, they immediately lead to the desired form of Eq. \eqref{eq23}, which, to high precision, accounts for both inverse decays and two-particle annihilations.

\bibliographystyle{apsrev4-1.bst}
\bibliography{CLANOK.bib}

\end{document}